\documentclass{aa}

\usepackage[varg]{txfonts}
\graphicspath{{./}{figures/}}
\usepackage{hyperref,graphicx}
\usepackage{makecell}

\begin{document}

\title{The formation and structure of iron-dominated planetesimals}
\author{Terry-Ann Suer \inst{1,2} \and Edgar S. Steenstra \inst{3,4} \and Simone Marchi \inst{5} \and John A. Tarduno \inst{1, 6} \and Ilaria Pascucci \inst{7} }
\institute{Laboratory for Laser Energetics, University of Rochester, Rochester, NY 14623, USA \and Steward Observatory, University of Arizona, Tucson, AZ 85745, USA \and Institute of Mineralogy, University of Münster, Münster, Germany \and Faculty of Aerospace Engineering, TU Delft, Delft, the Netherlands \and Southwest Research Institute, Boulder, CO, U.S.A. \and Department of Earth and Environmental Science, University of Rochester, Rochester, NY 14623, USA \and Lunar and Planetary Laboratory, University of Arizona, Tucson, AZ 85745, USA}

\abstract  
   {Metal-rich asteroids and iron meteorites are considered core remnants of differentiated planetesimals and/or products of oxygen-depleted accretion.}
   {Investigating the origins of iron-rich planetesimals could provide key insights into planet formation mechanisms.}
   {Using differentiation models, we evaluate the interior structure and composition of representative-sized planetesimals ($\sim$200 km diameter), while varying oxygen fugacity and initial bulk meteoritic composition.}
   {Under the oxygen-poor conditions that likely existed early in the inner regions of the Solar System and other protoplanetary disks, core fractions remain relatively consistent across a range of bulk compositions (CI, H, EH, and CBa). 
Some of these cores could incorporate significant amounts of silicon ($10-30$ weight \%) and explain the metal fractions of Fe-rich bodies in the absence of mantle stripping. Conversely, planetesimals forming under more oxidizing conditions, such as beyond snow lines, could exhibit smaller cores, enriched in carbon, sulfur ($>$ 1 weight \%), and oxides. Sulfur-rich cores, like those formed from EH and H bulk compositions, could remain partly molten, sustain dynamos, and even drive sulfur-rich volcanism. Additionally, bodies with high carbon contents, such as CI compositions, can form graphitic outer layers.}
   {These variations highlight the importance of initial formation conditions in shaping planetesimal structures. Future missions, such as NASA's Psyche mission, offer an opportunity to measure the relative abundances of key elements (Fe, Ni, Si, and S) necessary to distinguish among formation scenarios and structure models for Fe-rich and reduced planetesimals.}


    \maketitle

\vspace{-1.2em} 
\noindent\textit{Accepted for publication in \textbf{Astronomy \& Astrophysics}.}\\
\textit{DOI:} \href{https://doi.org/10.1051/0004-6361/202554674}{10.1051/0004-6361/202554674}

\section{Introduction} \label{sec:intro}

Iron (Fe) is a major planet-forming element and is likely present in significant proportions in all rocky planets in the Solar System. The Fe budgets of Earth, Mars, and Venus cluster around $25-35 \%$ by mass ($12-15 \%$ by volume) \citep{wanke1981}, while Mercury stands out with $67 \%$ by mass ($40 \%$ by volume) \citep{prentice2003}. Some asteroids, most notably M-type Psyche, are hypothesized to have Fe volume fractions ranging from $30 - 60 \%$ \citep{elkins2022distinguishing,dibb2024post}. The meteoritic record also contains hundreds of Fe meteorites, classified into 12 groups, which are predominantly believed to represent the disrupted cores of early planetesimals \citep{scott1975}. However, the formation of these and other putative metal-rich parent bodies could have been influenced by multiple factors, including disk chemistry and dynamic evolution \citep{johansen2022,bogdan2023curie}. Compounding the abundance of Fe in the meteoritic record is the relative scarcity of silicates, an observation referred to as the ``missing mantle paradox''\citep{vaci2021olivine}. While mantle-stripping events can explain aspects of some Fe-rich bodies \citep{asphaug2014}, they alone cannot fully account for the formation of high-density exoplanets, including super-Mercuries, thought to be predominantly composed of Fe ($\geq$ 60 \% by mass) \citep{barros2022hd,cambioni2025can}. Exploring accretion pathways for Fe-rich planetary building blocks could, therefore, add valuable insights into the broader debate surrounding planet formation. 

In the absence of significant fragmentation, the composition and structure of Fe-rich bodies can be constrained by examining major compositional end-members across varying redox conditions. The early Solar System, like other protoplanetary disks, exhibited substantial spatial and temporal compositional diversity, reflecting a complex and evolving redox landscape \citep{krot2000meteoritical,wadhwa2008, grossman2008redox}. Disk models suggest that C/O in the gas within the snow line can vary over time, which implies that oxidation conditions can change \citep[e.g.,][]{mah2023close} rather than remain fixed. Meteoritic evidence suggests that the earliest planetesimals formed under diverse redox conditions. Remarkably, both non-carbonaceous (NC) and carbonaceous (CC) iron meteorite parent bodies, representing contrasting redox states and disk environments, accreted almost contemporaneously, within the first Myr of Solar System history ($<0.5$ Myr for NC and $<1$ Myr for CC; \citep{kruijer2017age}. Subsequently, Jupiter's formation ($1 - 4$ Ma after CAI) is thought to have separated the NC and CC reservoirs \citep{kruijer2017age,desch2018effect} by inhibiting the flow of icy pebbles to the inner disk \citep[e.g.,][]{mah2024mind}. Isotopic and dynamical studies suggest that some later mixing occurred between these two reservoirs ( $\sim 5-11$ Ma after CAI), possibly due to scattering of material from the inner disk \citep{bermingham2020nc,zhang2022cores, raymond2017,bottke2005fossilized}. The building blocks of the asteroid belt could therefore have incorporated a range of compositions and redox states. 

Oxygen fugacity ($f\text{O}_2$) in both the gas and solid phases plays an important role in determining planetary compositions. 
The oxygen partial pressure of the gas determines the composition of equilibrium condensates, while the oxygen distribution among solid phases governs the composition of mineral assemblages \citep{rubin2003}. Several recent studies have shown that redox state determines the structure of planets by controlling the distribution of metals and silicates and the elemental distribution among these two reservoirs \citep{siebert2011,steenstra2020a,steenstra2020b}. Under reducing (relatively oxygen-poor) conditions, bodies can have higher metal contents, whereas oxidized bodies have oxygen available to form mantle oxides. The effects of variable $f\text{O}_2$ and highly reduced initial bulk compositions on potential metal-rich planetesimals have only recently been explored \citep[e.g.,][]{grewal2024accretion,spitzer2025comparison}.

The formation of Psyche and other M-type asteroids could potentially provide insights into Fe-rich and/or reduced (oxygen-poor) building blocks \citep{fornasier2010spectroscopic}. Most models of Psyche's formation describe it as a planetesimal core remnant \citep{cambioni2022}, and recent work explores the effects of the oxidized bulk composition on the parent body structure \citep{bercovici2022}. However, given uncertainties in composition and evidence for scattering of planetesimals from the inner disk, there is no reason to exclude more reduced initial scenarios. In addition, highly reduced materials have also been proposed to partially explain dense planetary bodies such as Mercury, Mercury-like exoplanets, and their building blocks \citep{cartier2019role,cioria2024mantle,mah2023forming}. 

Here, we explore the effects of $f\text{O}_2$ and bulk composition on the compositional stratification and structure of planetary building blocks, with an emphasis on reduced initial compositions and conditions. Our predictions for interior structure, surface composition, and densities can be compared with geochemical and geophysical measurements from future exploratory missions such as Psyche \citep{elkins2022distinguishing}. Assessing the formation of Fe-rich or reduced bodies in the context of recent disk evolution studies could offer valuable insights into the history of our Solar System and help explain the observed diversity of planetary compositions.

\section{Methods} \label{sec:methods}
We modeled possible interior compositions for $\sim$200 km  diameter bodies, representative of the minimum initial planetesimal mass \citep{polak2023high} and similar in size to Psyche. Due to limited constraints on the composition of Fe-rich and reduced building blocks, we evaluated a range of potential initial compositions based on primitive meteorites (CI, H, CBa, and EH; Table 1). Our objective was to assess the properties of bodies formed from materials with varying refractory and light element content, as well as differing initial redox states. The CI chondrites represent some of the most oxidized and volatile-rich meteorites, whereas CBa chondrites are among the most reduced and volatile-poor \citep{krot2002cr}. The H chondrites formed under intermediate redox conditions \citep{mcsween1993oxidation}, while the EH chondrites formed under highly reduced but S-rich conditions \citep{trieloff2022evolution}.  While these compositions are used to explore different scenarios, the histories of individual parent bodies lie outside the scope of the current study. However, recent studies have used similar compositions to study the mineralogy of reduced sub-Earths and exo-Mercuries \citep[e.g.,][]{steenstra2020c,cioria2024mantle}. In addition to the four end-members, we also considered mixing between CI and CBa in ratios of 1:1, 1:3, and 3:1 to explore the outcome of possible reservoir mixing between reduced and oxidized materials. The Fe content of the starting compositions ranged from 18.5 (CI) to 67 (CBa) wt.\% with significant variation in light elements. The Fe/Si ratio, an indicator of core-to-mantle proportions, also varies between 1.6 (H) and 9.6 (CBa) (Table 1).

\begin{table}
\tiny
\label{table:1}
\caption{Initial bulk compositions, major elements, densities, and redox ranges for the cases considered in this work.\protect\footnotemark{}
}
\begin{tabular}{c c c c c c c c c c|}
\hline
\makecell {Bulk \\ Comp} & \makecell {Mg \\ wt \%} & \makecell{Si \\ wt \%} & \makecell{Fe \\wt \%} & \makecell{S \\wt \%} & \makecell{C \\wt \%} & \makecell{Ni \\wt \%} & \makecell {Grain \\ $\rho$ \\ g/cm$^3$} & \makecell{Redox\\ ($\Delta\text{IW}$)} \\
\hline
CB & 5.6 & 7.0 & 67.3 &0.5&0.78&4.75& 5.52 & $-2$ \\ 
EH& 10.6 & 16.7 & 29.0 & 2 & 0.4 & 1.75 & 3.62 & $-4$ to $-6$ \\  
H& 14.0 & 16.9 & 27.5 & 2.0&0.22 & 1.60& 3.88 & $+1$ to $+2$ \\
CI& 9.54 & 10.7 & 18.5 & 5.35 & 3.65 & 1.06 & 2.54 & $+4$ to $+7$ \\
1 CI:3 CB &6.59&7.93&55.1&1.71&1.50&3.83&4.78 & \\
1 CI:1 CB &7.57&8.85&42.9&2.93&2.22&2.91&4.03 & \\
3 CI:1 CB &8.56&9.78&30.7&4.14&2.93&1.98&3.29 & \\
\hline
\end{tabular}
\end{table}

\footnotetext{Compositions: CB \citep{rubin2003}; CH, CI, H, and EH \citep{alexander2019}. The estimate of 0.22 wt. \% C is used for H based on the C content of relatively unmetamorphosed Group 3 H chondrites \citep{moore1967}. The grain densities were determined from \cite{britt2002, consolmagno2008, macke2011density}. The redox states of the end members are \cite{rubin2003,campbell2002siderophile,mcsween1993oxidation,krot2002cr,mccoy1999partial,righter2016redox}}

Differentiation of the hypothetical parent bodies was simulated using an internally consistent chemical equilibrium differentiation approach in which the FeO content and core mass vary consistently with $f\text{O}_2$ \citep{palme2014,steenstra2016}. Using a mass balance, the concentrations of the elements of interest ($c_i$) were then calculated in both the core and mantle for each end-member and mixed composition: 

\begin{equation}\label{eqn:eqn1}
c^\text{c}_{i} = \frac{c^\text{BPB}_{i}}{M^\text{c}} + \frac{(1-M^\text{c})}{D^\text{met-sil}_{i}}\\
\end{equation}

\begin{equation}\label{eqn:eqn2}
c^\text{m}_{i} = \frac{c^\text{BPB}_{i}}{M^\text{m}} + (1-M^\text{{m}})*D^\text{met-sil}_{i}\\ .
\end{equation}

Here, $c^\text{c}_{i}$ is the concentration of element \textit{i} in the parent body core, $c^\text{m}_{i}$ is the concentration of element \textit{i} in the parent body mantle, $c^\text{BPB}_{i}$ is the concentration by weight of element $i$ in the bulk parent body (BPB), $M^\text{c}$ and $M^\text{m}$ are the parent body core and mantle mass fraction, and $D^\text{met-sil}_{i}$ is the metal–silicate partition coefficient for element \textit{i} defined as 

\begin{equation}\label{eqn:eqn3}
D^\text{met-sil}_{i} = \frac{[X]^\text{metal}_{i}}{[X]^\text{silicate}_{i}}\\,
\end{equation}

where ${[X]^\text{metal}_{i}}$ is the mole fraction of element \textit{i} in the metal and ${[X]^\text{silicate}_{i}}$ is the mole fraction of element \textit{i} in the silicate melt. 

The oxygen fugacity, $f\text{O}_2$, a variable that strongly affects $D^\text{met-sil}_{i}$, was defined relative to the iron-w{\"u}stite ($\Delta\text{IW}$) redox buffer. This buffer represents the equilibrium reaction between metallic iron and iron oxide (FeO) in the presence of oxygen \citep{o1993thermodynamic}. Expressed in logarithmic units, $\Delta\text{IW}$ is often used to quantify deviations from the above equilibrium in studies on metallic Fe core formation:

\begin{equation}\label{eqn:eqn4}
\Delta \text{IW} = 2 log \left(\frac{a^\text{silicate}_\text{FeO}}{a^\text{metal}_\text{Fe}}\right) = 2 log \left(\frac{X^\text{silicate}_\text{FeO}}{X^\text{metal}_\text{Fe}}\right)+ 2 log \left(\frac{\gamma^\text{silicate}_\text{FeO}}{\gamma^\text{metal}_\text{Fe}} \right) \\,
\end{equation}
where $a^\text{silicate}_\text{FeO}$ and $X^\text{silicate}_\text{FeO}$ are the activity and molar fraction of FeO in the silicate melt and $a^\text{metal}_\text{Fe}$ and $X^\text{metal}_\text{Fe}$ are the activity and molar fraction of Fe in the metal phase, respectively \citep[e.g.,][]{siebert2011}. The activity coefficient of Fe, $\gamma^\text{metal}_\text{Fe}$, was assumed to be constant throughout the range of the model. For most oxidized conditions, $f\text{O}_2$ is expected to be slightly overestimated. Silicon in liquid Fe has minor effects on $\gamma^\text{metal}_\text{Fe}$ ($\sim$0.5 to 0.7) \citep{steenstra2020a, righter2019effect, kilburn1997metal}.  Carbon may lower $\gamma^\text{metal}_\text{Fe}$ to 0.7 at graphite saturation, while S may increase $\gamma^\text{metal}_\text{Fe}$ to $\sim$3 for near-stoichiometric FeS liquids \citep{jennings2021}.  Metal-silicate partition coefficients were calculated across a range of $f\text{O}_2$ values for each bulk composition, considering the core mass fraction (CMF) and corresponding FeO mantle contents. The CMFs were calculated using the method summarized in \cite{palme2014}, equation 7.

The distributions of major elements between the core and mantle fractions were calculated using experimentally derived predictive metal-silicate partitioning expressions for C, S, Si, and Ni \citep{boujibar2014,steenstra2016, chi2014, Fischer2015, steenstra2020a, dalou2024review, suer2023distribution}. The variation of $D^\text{met-sil}$ for the elements considered here is shown in Figure A.1 of the Appendix. The parametrization for Ni was not calibrated for highly reduced conditions \citep{steenstra2016}; therefore, $D^\text{met-sil}_\text{Ni}$ was assumed to be constant at $\Delta\text{IW} \leq -4.25$, consistent with \citet{steenstra2020a}. While Ni is predicted to become more siderophile under reducing conditions, the presence of Si in the metal reduces its affinity for metal \citep{steenstra2020a}. Carbon solubility, defined as the carbon concentration at graphite saturation (CCGS) in Fe-Si alloys, was modeled following \cite{steenstra2020a} for highly reduced and low-S conditions. The model of \cite{steenstra2018} was used for more oxidized conditions and to account for the effect of S on C solubility in Fe alloys. The S solubility capacity of mantle-forming silicate melts and the S content at sulfide saturation (SCSS) were estimated using the model of \cite{namur2016} across the $f\text{O}_2$ range in this study, although the model was calibrated for highly reduced, low-S conditions. For more oxidized conditions, we applied the model of \cite{steenstra2018} to account for the effect of S on C in Fe-rich alloys.

Partition coefficients and solubilities were calculated assuming a constant core formation temperature of 1900 K and a pressure of 0.1 GPa, corresponding to the core-mantle boundary of a $\sim$200 km diameter rocky body. The partition coefficients of C and S are only weakly affected by temperature within a realistic range for asteroidal differentiation (i.e., 1600 to 1900 K). However, $D^\text{met-sil}_\text{Si}$ depends more strongly on temperature \citep{ricolleau2011,Fischer2015} and could decrease by 1.35 over this temperature range. The modeled Si content of the parent bodies equilibrated at 1900 K should therefore be considered as an upper limit. Finally, the ratio of non-bridging oxygen atoms to tetrahedral cations (NBO/T), a proxy for silicate melt composition and polymerization required for calculating log $D^\text{met-sil}_\text{C}$  and $D^\text{met-sil}_\text{Ni}$, was assumed to be 2.55. This value is applicable to primitive silicate melts present during full or partial melting of primitive chondritic bodies \citep{jana1997}.

The densities of the starting materials, summarized in Table 1, ranged from $2.50$ to $5.54 g/cm^{3}$ \citep{britt2002, macke2011density,consolmagno1998}. Grain densities (corrected for non-porosity) were used as proxies for the densities of the initial bodies in lieu of bulk densities that account for porosity. Densities for the modeled cores were approximated using an ideal linear mixing of the elemental components as a function of $f\text{O}_2$. Mantle densities were estimated using a mass-balance model.

We implemented a simple scheme to estimate the first-order effects of mantle removal on the CMF and bulk density. This mass-balance model did not account for impact-specific parameters such as energy, impactor size, velocity, or angle \citep{marcus2009collisional}. Instead, we explored changes in planetary structure by systematically removing a fraction of the silicate mantle. The new core mass fraction was then given by 

\begin{equation}\label{eqn:eqn5}
CMF_\text{new} = \frac{CMF_{0}}{CMF_{0} + (1 - CMF_{0})(1 - f)}.
\end{equation}

Assuming a two-component body, the bulk density was calculated as the average of the core and mantle densities,

\begin{equation}\label{eqn:eqn6}
\rho_{new} = (\frac{CMF_{0}}{\rho_\text{core}} + \frac{1-CMF}{\rho_\text{mantle}})^{-1}. 
\end{equation}

The model does not consider further chemical evolution beyond initial formation. 

\section{Results} \label{sec:results}
\subsection{Relative core fractions} \label{subsec:results1} 
\begin{figure}[htb!]
    \centering
    \includegraphics[width=.5\textwidth, height=.3\textheight, keepaspectratio]{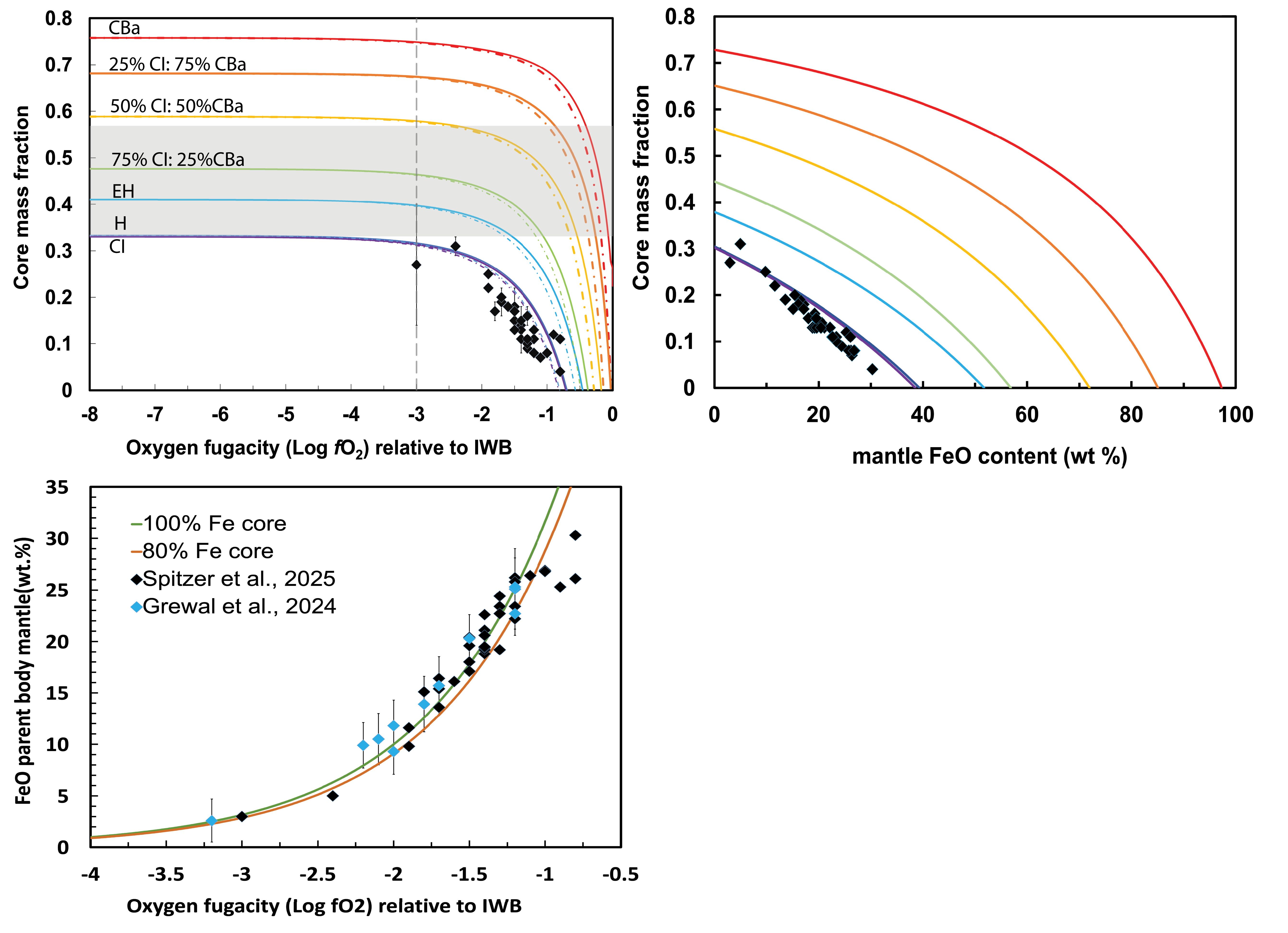}
    \caption{Top left: Modeled parent body core mass fractions (CMF) versus oxygen fugacity ($f\text{O}_2$) for the compositional cases considered here. The shaded horizontal bar represents CMF estimates for Psyche \citep{elkins2022distinguishing}. The dashed vertical line indicates the redox below which metallic Fe dominates. Top right: CMF evolution with mantle FeO content for our cases. Bottom: FeO versus $f\text{O}_2$ from our results. Data from recent studies on iron meteorites are overplotted for comparison \citep{grewal2024accretion,spitzer2025comparison,hilton2022chemical}}
    \label{fig:fig1}
\end{figure}

 We find that the CMF varies only slightly at $f\text{O}_2$ levels more reducing than $\Delta\text{IW}-3$, across the range of compositions considered (Fig. \ref{fig:fig1}, top left) for the pressure and temperature of the models. The maximum CMF varied from 0.32 for H and CI to 0.75 for CBa under the most reducing conditions, largely reflecting the initial Fe to Si ratios. Iron favors Fe$^{0}$ metal (low oxidation state) below $\Delta\text{IW} -3$, leading to near-invariant CMFs under these reducing conditions. The CMF for all compositions was lowered as $f\text{O}_2$ increased above $\Delta\text{IW} = -2$ as more Fe oxidized to FeO (Fig. \ref{fig:fig1}, top right). The CMF for all compositions becomes negligible (that is, no metallic Fe core) under conditions more oxidized than $\Delta\text{IW} -1$ when FeO (rather than Fe), S, and C become significant core components \citep[e.g.,][]{bercovici2022, corgne2008c, buono2011fe}. Overall, these models indicate that Fe-rich cores could form across a broad range of redox states, bulk compositions, and thermodynamic conditions.
 
 These results imply that the cores or metal mass fractions of planetesimals that accreted under the most reduced conditions can be directly linked to bulk composition. However, caution is warranted, as Fe-rich bulk compositions (e.g., CBa) can also form high CMF bodies under oxidizing conditions ($\Delta\text{IW} = -1$). Additionally, bodies with relatively low CMFs ($\sim$ 0.3) could also form at highly reduced conditions from H- or CI-like end-members. Considering current estimates of porosities for metal-rich asteroids, a CBa-rich bulk composition could yield a body with a core fraction of 60 \%, the upper limit of Psyche's estimated metal fraction \citep{elkins2020observations}. In contrast, EH-rich material could produce a body with a maximum CMF of $\sim$ $40 \%$ (Fig. \ref{fig:fig1} (top-left)).

\subsection{Core compositions} 
\label{subsec:results2}
\begin{figure}
    \centering
    \includegraphics[width=0.5\textwidth, height=0.45\textheight]{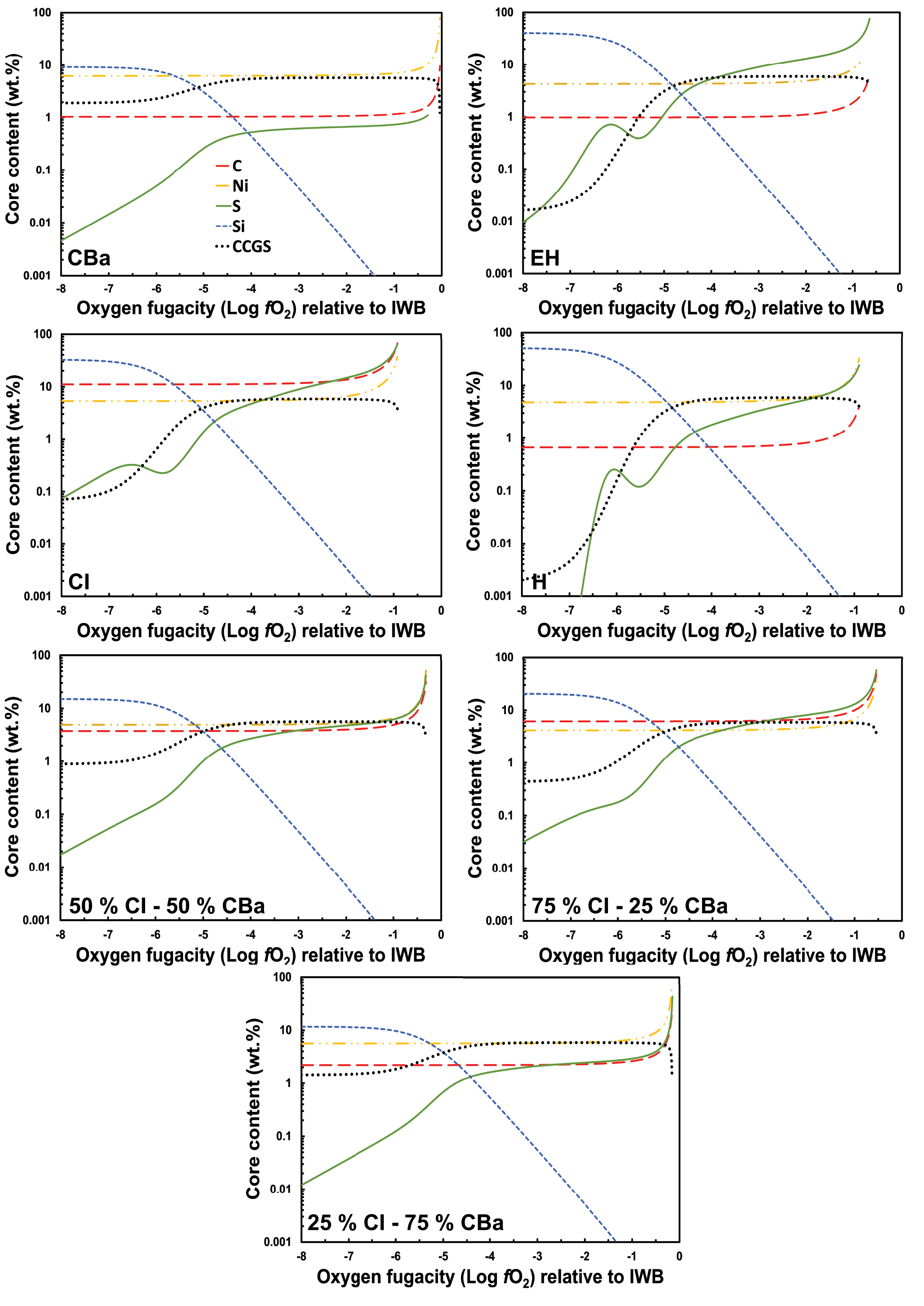}
    \caption{Modeled core light element compositions versus oxygen fugacity ($f\text{O}_2$) for the  compositions considered. Metal-silicate partition coefficients used in the models are from \citep{boujibar2014,steenstra2016, chi2014, Fischer2015, steenstra2020a}. The  carbon concentration at graphite saturation (CCGS) for the metal fractions of each bulk composition is plotted for comparison \citep{steenstra2020a,steenstra2018}.}
    \label{fig:fig2}
\end{figure}

Core composition is determined by the competing thermodynamics of elemental partitioning and solubility. Although core fractions remain stable, core light element composition changes across the redox range of the study (Fig. \ref{fig:fig2}). Due to its siderophile behavior, Si is the dominant light alloying element under the most reduced conditions ($\Delta\text{IW}<-6.5$). Its initial core abundance is directly reflected by the amount of Si in the parent body building blocks, with EH and H bodies containing significantly more Si in their cores. Between $\Delta\text{IW} = -5$ and $-4$, Si core content declines across all bulk compositions, driven by a decrease in $D^\text{met-sil}_\text{Si}$. Under more oxidized conditions, both S and C are partitioned into Fe in appreciable amounts, with S being equally or more abundant than C in the cores of EH bodies. The carbon saturation limits of Fe-rich liquids increase with oxidation (indicated by the dotted black lines in Fig. \ref{fig:fig2}) leading to graphite saturation in most moderate to highly reduced cases, except for a pure bulk CBa body. Nickel, a strongly siderophile element, primarily partitions into the core, resulting in near-constant Ni concentrations over most model conditions. Consequently, core Ni contents largely reflect the Ni content of the assumed building blocks, adjusted for the CMF. However, under the most oxidized conditions, the relative core abundance of Ni, along with C and S, increases significantly, in agreement with previous findings \citep{bercovici2022}. 

Extensive carbon loss during parent body metamorphism likely prevented the preservation of C-rich or graphite-saturated cores in the meteorite record, even if such compositions are thermodynamically permissible \citep[e.g.,][]{grewal2022internal, grewal2025tracing}.

\subsection{Mantle compositions}
\label{subsec:results3}
\begin{figure}
    \centering
    \includegraphics[width=0.5\textwidth, height=0.45\textheight]{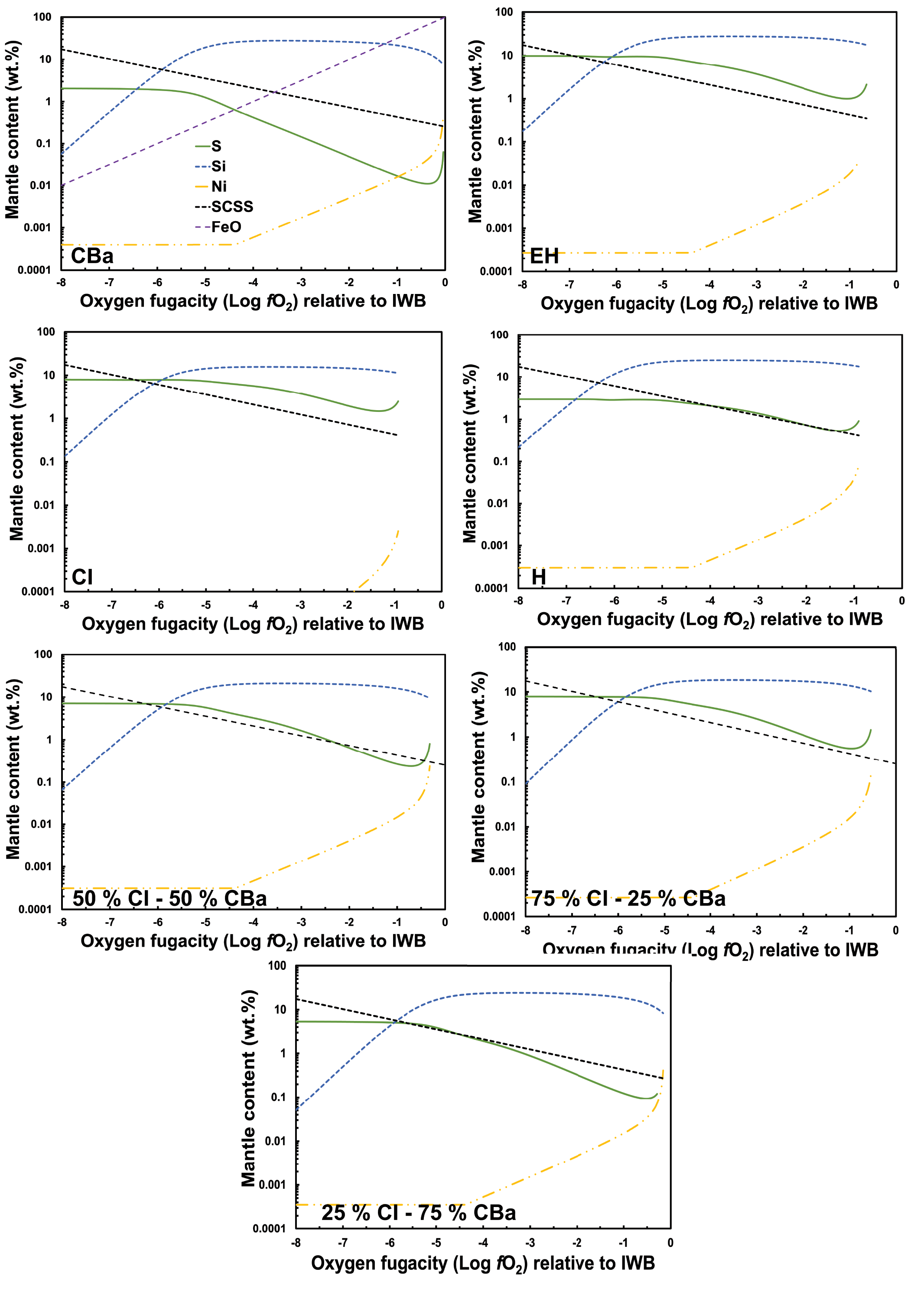}
    \caption{Modeled minor elemental mantle compositions for the different initial bulk compositions as a function of $f\text{O}_2$. Compositions are obtained from a core-mantle differentiation model with experimentally determined metal-silicate partition coefficients \citep{boujibar2014,steenstra2016, chi2014, Fischer2015, steenstra2020a}. The estimated SCSS is plotted for comparison \cite{namur2016}. The FeO evolution path shown for CBa represents the trend for other compositions.} 
    \label{fig:fig3}
\end{figure}

\begin{figure}[htb!]
    \centering
    \includegraphics[width=0.5\textwidth, height=0.5\textheight, keepaspectratio]{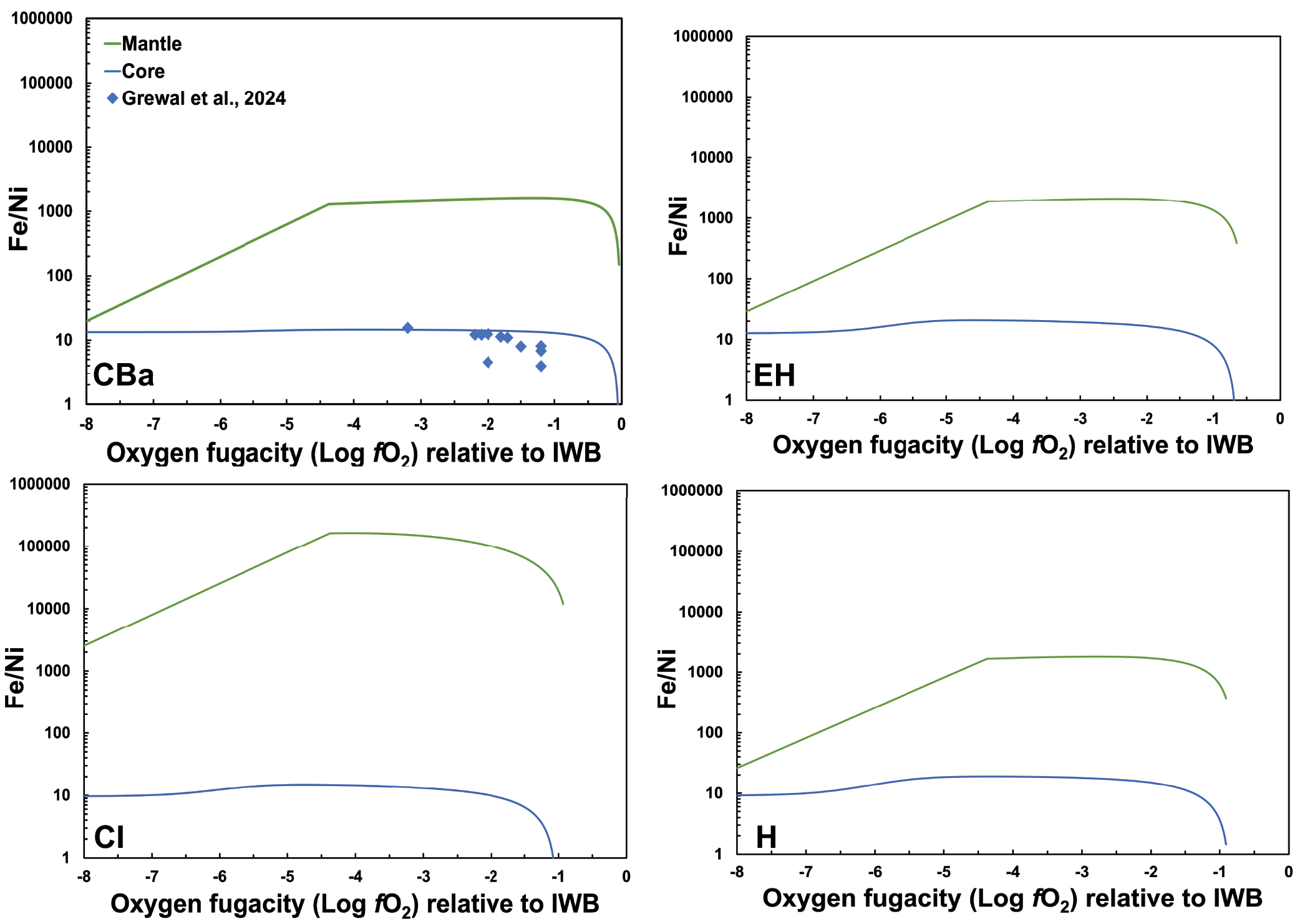}
    \caption{Summary of the Fe/Ni ratios for the core and mantle fractions of the four pure end-member bulk compositions versus $f\text{O}_2$. The Fe/Ni ratios of Fe meteorites from \citet{grewal2024accretion} are overlaid in panel 1. For the cores of Earth and Mars, Fe/Ni ratios range from 16 to 20  \citep{dreibus1996cosmochemical,khan2018geophysical}, and from hundreds to thousands in rocky mantles and differentiated stony meteorites\citep{palme2014}.}
    \label{fig:fig4}
\end{figure}

The chemistry of planetesimal mantles is strongly affected by redox conditions. As oxygen becomes more available, Fe becomes oxidized, increasing the FeO content of silicates (Fig.  \ref{fig:fig1}, lower panel). Recent studies of equilibrium mineral assemblages for a variety of bulk compositions indicate that the mantles of reduced bodies are dominated by reduced silicates such as enstatite and pyroxenes \citep{cioria2024mantle}, whereas more oxidized mantles contain olivine and orthopyroxene \citep{toplis2013chondritic}.  Our results suggest that other major element concentrations (including Ni, Si, S, and C) also vary with redox (see Fig.  \ref{fig:fig3} and Fig. \ref{fig:fig4}). Below approximately $\Delta\text{IW}-3$, under highly reducing conditions, Ni remains predominantly metallic (Ni$^{0}$) and partitions efficiently into the core, resulting in negligible concentrations in the mantle. At higher oxygen fugacities, particularly above $\Delta\text{IW}-1$ to $+1$, Ni becomes oxidized to Ni$^{2+}$ and can be incorporated into mantle silicates, where concentrations can reach thousands of ppm. Although our calculations suggest that most Si would partition into the core under highly reduced conditions, this would imply a Si-free mantle, an unrealistic scenario that contradicts the presence of FeO-free silicates in reduced achondrites \citep{britt2003}. This discrepancy arises from the non-ideal behavior of FeO in silicate melts at low $f\text{O}_2$. Experiments show that below $\Delta\text{IW} = -7$, FeO activities deviate significantly, increasing FeO contents in silicate melts under extreme reduction \citep{steenstra2020a}. 

The S and C contents of silicate melts depend on melt composition, P, T, and redox state \citep{O'neill2002,namur2016,smythe2017, steenstra2020b, chi2014}. Our results show that while S favors Fe cores at $f\text{O}_2 > \Delta\text{IW} = -4$, it is enriched in silicate melts under more reducing conditions. However, its high solubility in low-FeO silicate melts diminishes the effects of its non-siderophile behavior \citep{namur2016,steenstra2020b}. Our results further suggest that sulfide saturation at $-6 < \Delta\text{IW} -4$ is plausible for most compositions except CBa (Figure \ref{fig:fig3}), assuming no significant volatile-related S loss. Graphite saturation from cores may provide a reservoir of carbon that plays a role in core-mantle boundaries, mantles, and crusts. The low solubility of C in silicate mantles (tens of ppm; \citep{chi2014}) implies that C saturation could be widespread in the mantles of CI-rich planetesimals across all redox conditions considered, and in H and EH-rich bodies equilibrated under conditions more reducing than $\Delta\text{IW} \sim-5$ (Figure \ref{fig:fig3}). Although the lack of evidence in the literature for carbon-rich cores could be due to loss through devolatilization, their existence cannot be ruled out \citep{grewal2022internal,grewal2025tracing}

\subsection{Density and stratification} \label{subsec:results4}
\begin{figure}[hbt!]
    \centering
    \includegraphics[width=0.4\textwidth, height=0.4\textheight, keepaspectratio]{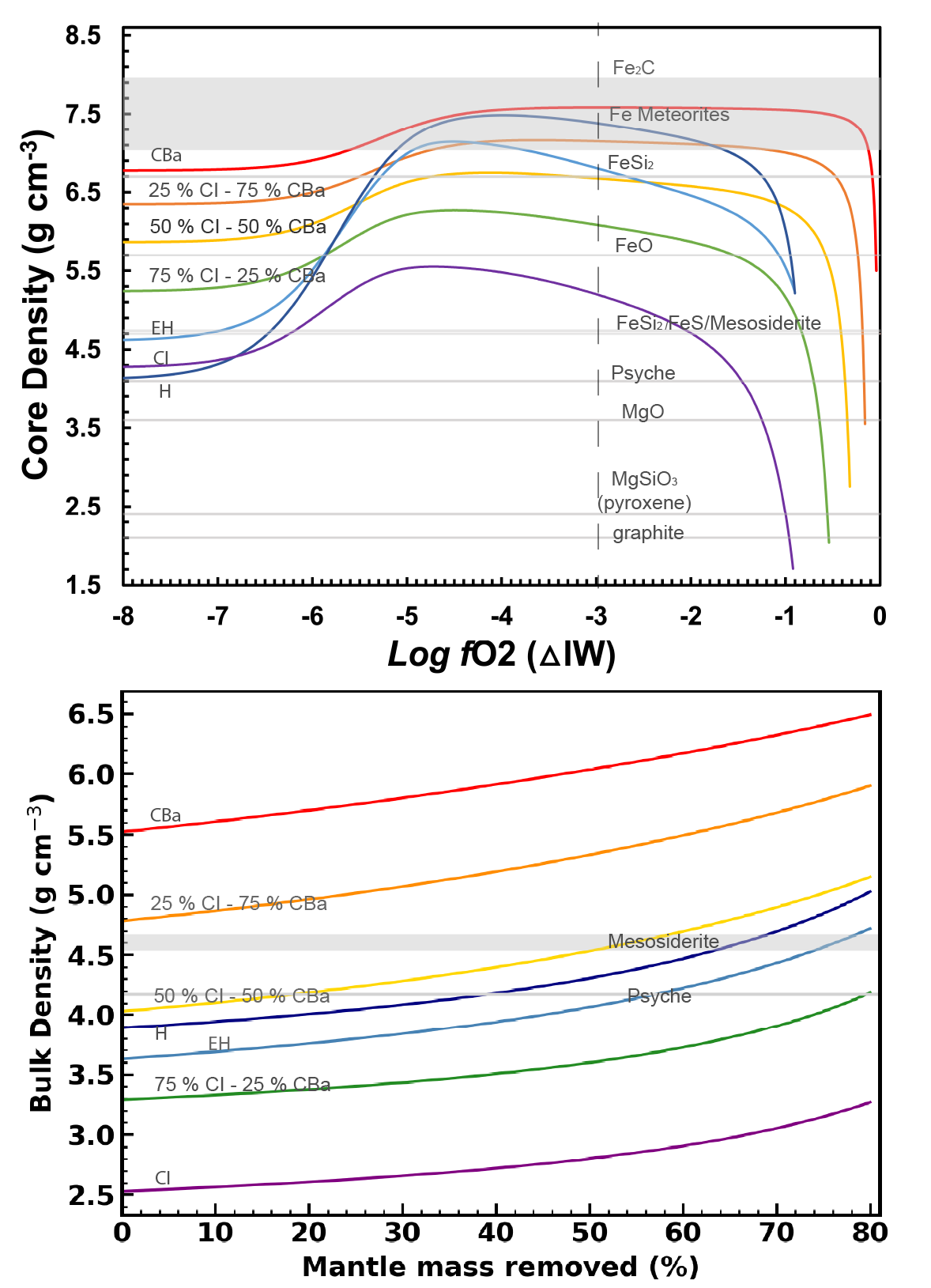}
    \caption{Top: Core densities vs. $f\text{O}_2$ for compositions considered in this study compared to densities of minerals, meteorites, and Psyche. Bottom: Change in bulk density for different amounts of mantle stripping. Densities are calculated by mass balance. Sources of initial densities: meteoritic \citep{britt2002, macke2011density, ostrowski2019physical}, Psyche \citep{farnocchia2024mass}, and minerals \cite{dziewonski1981preliminary, hauck2019mercury,zhu2021density}. A density of 2.5 $g/cm^{3}$ is the minimum plausible mantle density.}
    \label{fig:fig5}
\end{figure}

Planetary differentiation redistributes material based on density and chemical affinities, leading to internal stratification. Among the abundant planet-forming elements, Fe, which is the densest, segregates to form planetary cores. These core densities are governed by the solubility of lighter elements in Fe. Our calculated core densities are highest at moderately redox conditions, but decrease significantly at both more reduced and the most oxidized conditions. The range of core densities is 4.2 - 6.6 g/cm$^{3}$ at $\Delta\text{IW} = -8$ and $5.3 -7.5$ g/cm$^{3}$ at $-5 \geq \Delta\text{IW} \leq -3$ (Fig. \ref{fig:fig5}).  This observation is partly due to the crossover in partitioning of light elements Si and S in Fe at $\Delta\text{IW} \sim$$-5$ (see Appendix, Fig. A.1). Core densities are plotted over various phases that could plausibly occur in the cores or mantles of planetesimals. The minimum plausible mantle density based on previous studies is $\sim$2.2 g/cm$^{3}$ \citep{cioria2024mantle}. 

Density distribution, and consequently internal structure, is affected by the precipitation of light elements. In C-rich compositions, graphite efficiently segregates, floats in magmas, and can form a graphitic crust. The observation of a darkening agent on Mercury's surface suggests that this process could have occurred in even larger bodies \citep{steenstra2016,mccubbin2017,nittler2019surface}. In S-rich cases, sulfide bearing mantle phases from sulfide saturation \citep{anzures2020effect} can also lead to further stratification. Figure \ref{fig:fig6} (top panel) illustrates plausible cases for small bodies across the redox range of this study. Using mass-balance arguments, we also calculated the effect of removing fractions of the mantle on the bulk densities obtained at $\Delta\text{IW} = -3$ (Fig. \ref{fig:fig5} (low)). These calculations did not consider porosity, which would lower densities.

\section{Discussion} \label{sec:discussion}

\subsection{Compositional predictions for asteroid rendezvous missions} \label{sec:disc1}

Our results can be directly compared with future compositional measurements from asteroid rendezvous missions to help determine the structure and origins. The Multispectral Imager (MSI) on the Psyche mission can distinguish between metal and silicate components and is likely capable of identifying the different sulfide mineral groups \citep{elkins2022distinguishing, dibb2024post}. The Gamma Ray Neutron Spectrometer (GRNS) onboard Psyche can measure abundances of key elements such as Fe, Ni, Si, O, and C \citep{mccoy2022deciphering}. 
 
Bulk Ni content and Fe/Ni ratios of core fragments can help distinguish between NC and CC origins by constraining redox conditions at formation \citep[e.g.,][]{grewal2024accretion, spitzer2025comparison} (Fig. \ref{fig:fig4}). An instrument such as the GRNS, which can detect Si concentrations above 2.5 wt.\%, could reveal evidence of highly reduced accretion \citep{mccoy2022deciphering}. However, distinguishing indigenous Si from exogenous Si from later accretion can be challenging \citep{mccoy2022deciphering}. The GRNS also measures C content with a sensitivity of $1.4 \pm 0.9$ wt.$\%$ \citep{mccoy2022deciphering}, but our results show that this threshold is exceeded only under more oxidized conditions ($\Delta\text{IW} \geq -1$), except for CI-rich bulk compositions. Subsequently, metallic C contents in highly reduced cores cannot be accurately determined \citep{mccoy2022deciphering}. Carbon measurements may also indicate the presence of exsolved graphite (see section \ref{subsec:results3}, \ref{subsec:results4}. Sulfur-rich phases, detectable by either the GRNS or MSI, might be present on more oxidized bodies. Although S preferentially remains in silicate melts rather than cores, S-rich phases on the surface may also be linked to ferrovolcanism \citep{shepard2021}. 

 Oxide and silicate features may be detectable in the infrared (IR) spectrum of mantle fragments, although the Psyche spacecraft does not carry an IR instrument. Recent studies \citep[e.g.,][]{hardersen2011m,sanchez2016detection} analyzed pyroxene band depths in the near-IR spectra of M-type asteroids including Psyche, identifying absorption features consistent with low-Fe pyroxene or enstatite. These minerals are found in EH and achondrites and are generally indicative of reduced conditions \citep{mccoy2022deciphering, cioria2024mantle}. An IR instrument would therefore be a valuable addition to future asteroid rendezvous missions.  

 \begin{figure}[htb!]
    \centering
    \includegraphics[width=0.45\textwidth, height=0.45\textheight, keepaspectratio]{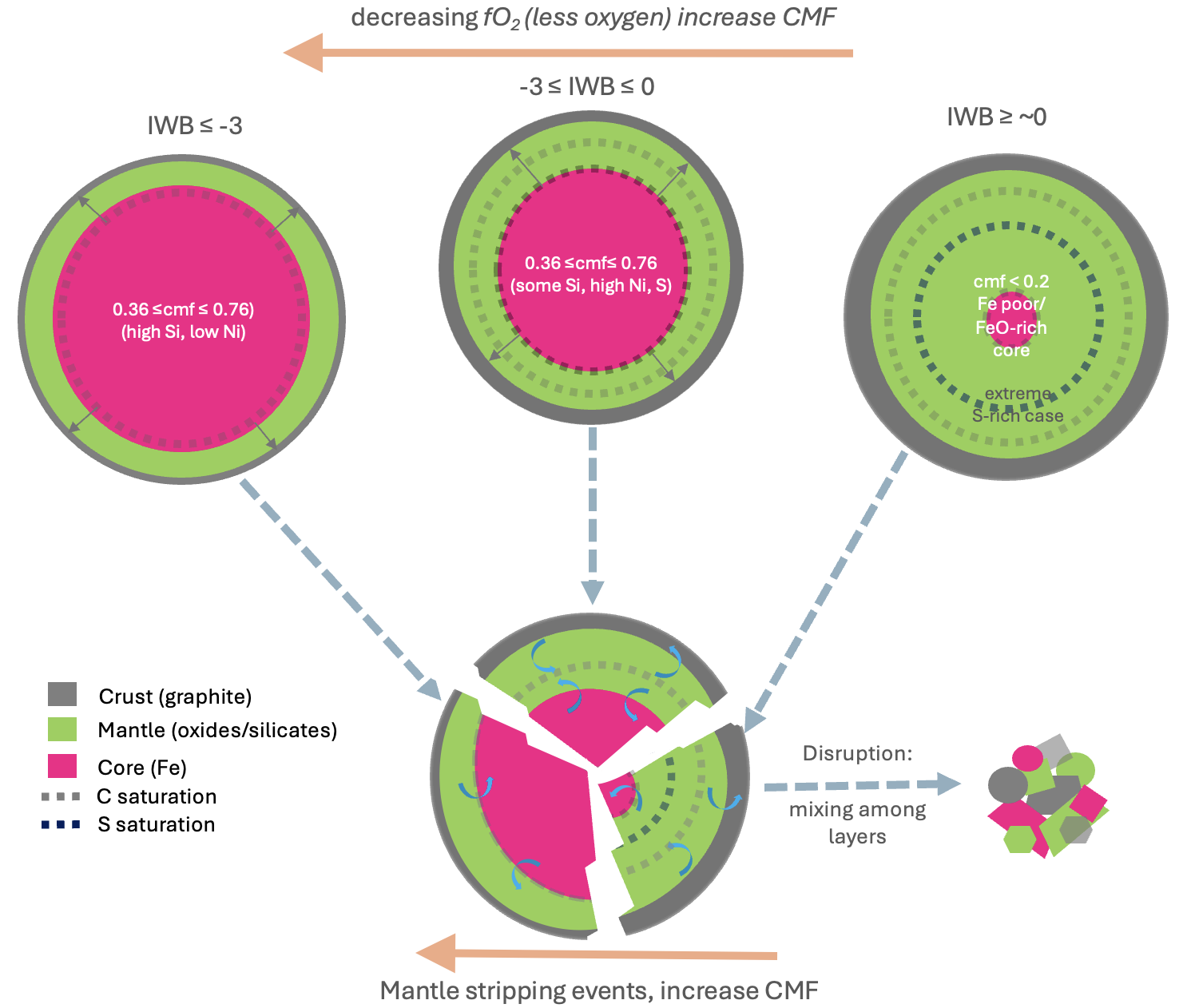}
    \caption{Plausible planetesimal structures resulting from a range of redox and disruption scenarios. Redox-sensitive results indicate that relatively small cores form for most bulk compositions above $\text{IW}$, but core sizes can change dramatically between -3 $\leq$ $\Delta\text{IW} \leq 0$. Disruption can produce additional scenarios that are either Fe- or silicate-rich, depending on the details of the events. Mantle stripping can lead to bodies with larger CMFs as well as bodies formed from mixed fragments \citep{asphaug2014}.}
    \label{fig:fig6}
\end{figure}

\subsection{Physical characteristics of reduced asteroids} \label{sec:disc2}

Our internal structure predictions are primarily based on composition and redox and can be tested with future geophysical measurements of Psyche and other asteroids \citep{zuber2022psyche}. Psyche's current bulk density falls in a cluster of CBa-like and mesosiderite bulk compositions, which represent an average of Fe and silicates (Fig. \ref{fig:fig5}) \citep{viikinkoski201816}. Before mantle removal, end-members containing significant CBa exceed current estimates of Psyche's bulk density (4.172 g/cm$^{3}$ \citep{farnocchia2024mass}) (Fig. \ref{fig:fig5}, lower panel. Lower-density end-members containing H and EH can approach Psyche's density at mantle loss fractions $\geq 40$ \% (Fig. \ref{fig:fig5}, lower panel; Fig. \ref{fig:fig7}). Pure CI could only explain Psyche's density after $\sim80 \%$ mantle removal, a plausible but unlikely scenario. Alternatively, a rubble pile of metal-rich compositions could also explain Psyche's density. Some scenarios are highlighted in Fig. \ref{fig:fig6}

Porosity also plays an important role in determining asteroid density and internal structure and can be modified by evolutionary processes such as mantle-stripping impacts \citep{Asphaug2017,zhang2023}. Porosities for the bulk compositions considered here range from 6 to 16 \% \citep{consolmagno1998, britt2002}, with asteroid estimates up to $30\pm 20\%$ \citep{britt2002}, consistent with a rubble pile of loosely bound fragments. Although mantle stripping cannot be ruled out, extreme stripping would require high degrees of parent body porosity to match the density of a body such as Psyche to a core fragment \citep{shepard2021}. Mantle stripping might also conflict with the proposed presence of hydrous phases on Psyche \citep{takir2016}, although exogenous hydrated silicates could also have been implanted later \citep{avdellidou2018exogenous}. Impact experiments on foundry alloys report extensive cracking, which could increase porosity in reduced or metal-rich bodies \citep{marchi2020hypervelocity}. 

If Psyche's metal fraction is lower than the estimated 30 to 60 vol\% \citep{elkins2022distinguishing}, possibly due to underestimated bulk porosity, most end-member bulk compositions used in this study can account for the required metal fraction, especially at conditions more reduced than $\Delta\text{IW} = -1$. However, if the bulk porosity of Psyche exceeds 40\% \citep{Flemingetal2022} and its metal fraction is higher, fewer bulk compositions remain plausible (Fig. \ref{fig:fig1}). Of the known meteorite groups, only CBa- or possibly a mesosiderite-like bulk composition could explain the metal content in the latter scenario. However, Psyche could still possess a large CMF or Fe content despite a relatively modest bulk density due to low self-gravity.

Our results also have implications for other physical observables. The dielectric constants of $10-30$ estimated for Psyche's surface by ALMA suggest a mixed metal-silicate surface \citep{dekleer2021,cambioni2022}. However, this range also overlaps with the dielectric constants of graphite and pyrite (FeS$_{2}$) \citep{lin2020,pereira2023}, which could indicate sulfide or graphite saturation or be linked to ferrovolcanism \citep{johnson2020,shepard2021}. Further observations are needed to confirm whether these processes occurred on Psyche.

\subsection{Dynamos in Fe-rich planetesimals}
\label{sec:disc3}
Our results offer compositional and structural models that can be used in simulations to evaluate the existence and longevity of core dynamos in Fe-rich planetesimals. Recent studies have explored the possibility of dynamos in small asteroidal bodies \citep[e.g.,][]{tarduno2012evidence, bryson2017paleomagnetic}. Evidence of magnetic fields has been found in some Fe meteorites that preserve remanent magnetization, with cooling-rate analyses suggesting that their parent bodies may have undergone either inward or outward solidification \citep{Scheinberg2016}. However, bulk Fe meteorites are generally poor magnetic recorders \citep{nagata1978magnetic} as they can retain intrinsic magnetization unrelated to past ambient magnetic fields \citep{brecher1977thermoremanence}. Therefore, if Psyche and other Fe-rich asteroids are primarily composed of metal, their ability to preserve ancient dynamo-generated fields may be limited. Instead, the presence of residual mantle or silicate crust, which can host fine-grained magnetic particles, would be more favorable for retaining a record of past dynamo activity.
 
 Models show that bodies with radii as small as 50 km could generate a dynamo for more than 10 Ma if the cores solidify outward \citep{nimmo2009energetics}. A dynamo may also be possible in the inward solidification case, with the outer shell preserving a magnetic signature, though initial mantle thickness has little impact on core dynamo evolution \citep{Scheinberg2016}. In addition, the presence of light elements S, C, and Si could play a key role by lowering the core liquidus temperature and increasing the likelihood of thermo-chemical convection in liquid Fe cores \citep{Breuer2015_core_review,neufeld2019top}. Recent work also suggests that a dynamo can be generated within the inner core of a rubble pile due to solidification and the expulsion of light elements at the core-mantle boundary \citep{zhang2023}. 
 
 The detection of a strong magnetic field on Psyche or another asteroid would indicate the presence of an early core dynamo and likely extensive igneous differentiation \citep{weiss2023psyche}. Metal fractions of such bodies could be Si-rich if the bodies accreted under highly reduced conditions, while S and/or C would be present if they accreted under more oxidized conditions (Fig. \ref{fig:fig3}). In the latter case, S and C would also likely be present as mantle sulfides or carbides, respectively. For a more oxidized body, among the possible light elements, S would be more efficient in decreasing the core liquidus. The Fe-FeS eutectic temperature is more than 500 degrees lower than the liquidus of pure Fe at 1 bar \citep{buono2011fe}, while C and Si decrease the liquidus temperature by $\sim$375 and $330\;^\circ\text{C}$, respectively \citep{massalski1986binary,buono2011fe}. Recent experiments show that in oxidized, sulfur-rich planetary materials, sulfide melts can percolate and fractionate before silicate melting, potentially forming metallic cores earlier and under cooler conditions than traditional models assume \citep{crossley2025percolative}. A thermo-chemically driven dynamo on an asteroidal body would therefore be most easily sustained in an FeS-rich system \citep{mcsween2002thermal,neumann2018modeling}. Generating and sustaining a dynamo is more probable under oxidized conditions where core S concentrations exceed a few weight percent—typically at $f\text{O}_2$ of $\Delta\text{IW} = -3$ or higher in most scenarios (Fig. \ref{fig:fig3}) \citep{williams2009bottom,zhang2023}. However, dynamo generation on more reduced EH-rich bodies cannot be ruled out.
 
\subsection {Reduced bodies in the meteoritic record}
\label{sec:disc4}

Although our results apply primarily to initial compositions and formation conditions, they can be compared to meteoritic compositions to infer conditions in the early Solar System. Several meteorite groups are believed to originate from reduced parent bodies: including enstatite (EH) chondrites, aubrites \citep{keil1989enstatite,steenstra2020c}, ureilites \citep{goodrich1992ureilites}, achondrite lodranites \citep{mittlefehldt1998non}, and winonaites \citep{benedix2000petrologic}. Most of these parent bodies formed within 2 Myr after CAIs and  experienced high temperatures and some igneous processes, although EH chondrites may have condensed directly from a reduced nebular gas \citep{krot2000meteoritical, grossman2008redox}. Many of the 60 Fe meteorite parent bodies, with sizes ranging from 50 to 600 km, are also inferred to have formed in an early reducing environment \citep{benedix2000petrologic, mccoy2022deciphering}, at $f\text{O}_2 < \Delta\text{IW}$.   
 
The Ni content in Fe meteorites is an indicator of initial redox \citep{grewal2024accretion, hilton2022chemical,spitzer2025comparison}. The Ni contents of CC Fe meteorites are generally $\geq$ 10 wt. \%, reflecting more oxidized conditions, whereas NC Fe meteorites typically contain 1 to 10 wt. \% Ni \citep{mccoy2022deciphering}. In our results, the Ni core contents range from 1 to 10 wt. \% across most of the redox range considered, consistent with NC families. However, under conditions more oxidized than $\Delta\text{IW} > -1.5$, Ni contents increase dramatically. Fig. \ref{fig:fig4} \citep{grewal2024accretion} shows a comparison of modeled Fe/Ni ratios with those of recent studies on Fe meteorites. Our results also predict a higher core S content in the cores of CI-rich, H, and EH compositions under oxidizing conditions, in contrast to the results of \citep{zhang2022cores} who report low core S content in the cores of CC bodies. Volatile loss could play a role in lowering initial S abundances \citep{hirschmann2021early}. 

Under reducing conditions, our results show that Si partitions strongly into Fe and would be expected in core fragments of reduced bodies. Although some Fe meteorites contain dissolved Si, it is also commonly found as oxides or silicates mixed with the metal. This is observed in Fe groups such as IIE and IAB and is often interpreted as evidence of impact-induced mixing between molten metal and silicates \citep{wasson2002iab,ruzicka2000geochemical}. Alternatively, Si initially dissolved in metal may have later exsolved or transformed into silica and other oxidized phases due to disruption or interactions with oxidized materials. Moreover, scattering and migration from the inner to outer regions of the disks could explain the late-stage oxidation inferred in certain Fe-rich meteorites \citep{bottke2006iron, hunt2022dissipation}. Some Fe meteorites contain unusually high amounts of Si in solution, indicative of highly reduced formation conditions. For example, the Tucson meteorite contains approximately 0.8 wt. \% of Si, while Horse Creek has 2.5 wt. \% \citep{miyake1974tucson,wai1969silicon}. The presence of silicates such as orthopyroxene and olivine in the Tucson meteorite further supports its formation under extremely reducing conditions \citep{miyake1974tucson, pack2011silicon}. 

Shock metamorphism is a key tool for studying terrestrial impact sites, but its signatures in Fe meteorites are harder to interpret \citep{scott1975}. Rapid or variable cooling rates in Fe meteorites are often linked to mantle stripping by energetic impacts or partially molten mantles \citep{goldstein2009iron}. For example, cooling rates for group IVA Fe meteorites range from 100 to 6,000 K/Myr, and increase with decreasing bulk Ni content \citet{yang2007iron}. The rapid cooling suggests a lack of insulation due to disruption, whereas the low Ni content implies reduced formation with a thin or highly conductive mantle, possibly containing graphite. Moreover, taenite cloudy microstructures indicate that subsequent impacts did not generate sufficient heat ( $T>$ $\sim$700 K) to erase these features \citep{einsle2018nanomagnetic}, suggesting that highly energetic impacts are not represented in Fe meteorites with cooling rate data. Simulations also indicate that hit-and-run collisions between planetesimals were too slow to induce significant melting  \citep{asphaug2010similar}. Gun-range experiments further support this, showing that impact melting is unlikely at main belt collision velocities \citep{alexander2022benchmarking}. If Fe meteorite parent bodies largely originate as reduced bodies with limited silicate mantles, this may partly explain the ``missing mantle paradox''\citep{vaci2021olivine,rider2025mystery}. 

\begin{figure}[t!]
    \centering
    \includegraphics[width=0.5\textwidth, height=0.5\textheight, keepaspectratio]{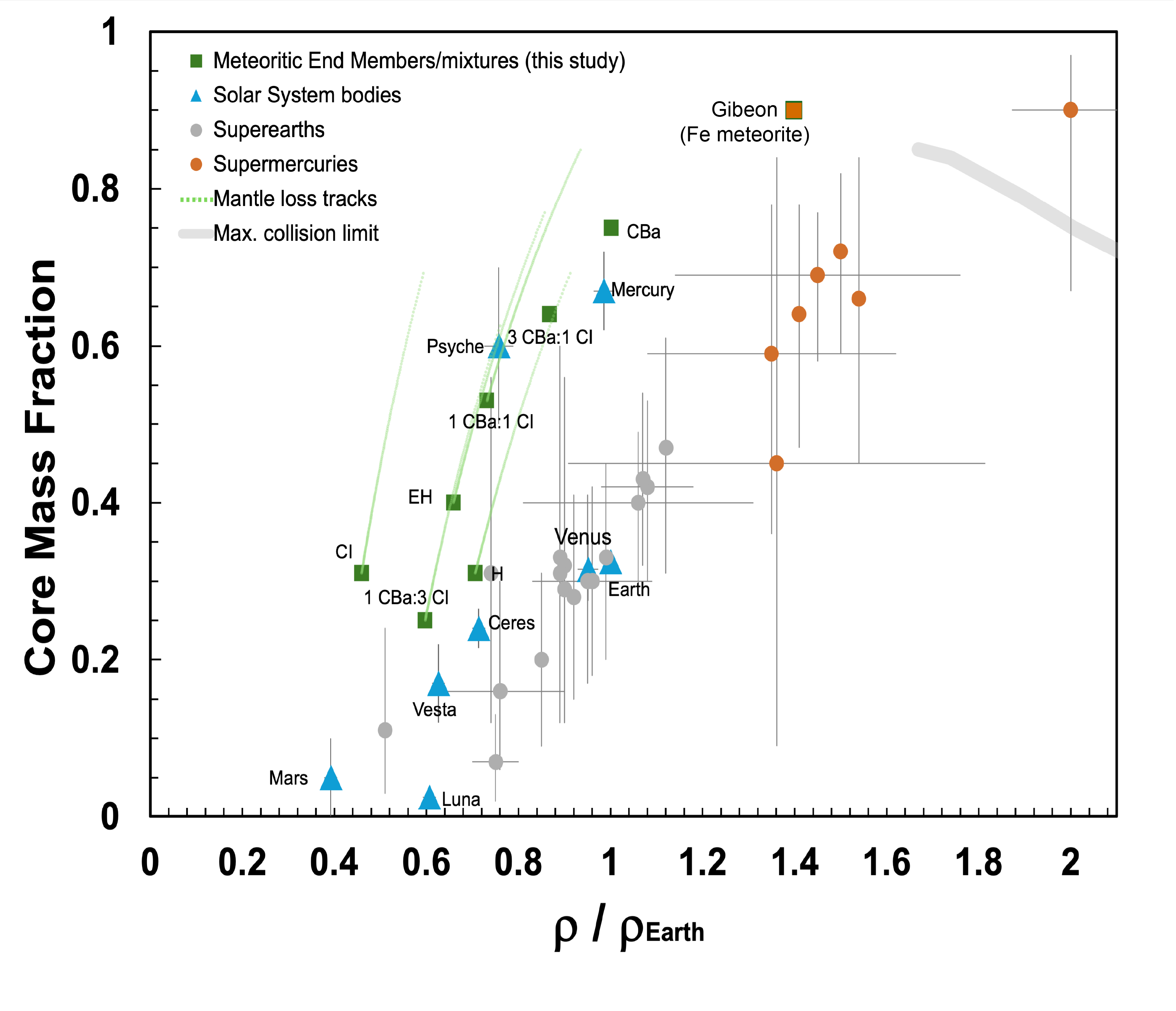}
    \caption{Core mass fractions vs. normalized density for a range of planetary bodies and meteorites. Exoplanet data are derived from recent studies \citep{adibekyan2021compositional,barros2022hd, goffo2023company} and the NASA Exoplanet Archive. Compositions for the Earth and other Solar System bodies and meteorites are from various sources \citep[e.g.,][]{grewal2024accretion,alexander2019, rubin2003, palme2014,ostrowski2019physical,wasson2001fractionation}. The maximum collisional limit is from \citep{marcus2010minimum}. \protect\footnotemark{}} 
     \label{fig:fig7}
\end{figure}

\footnotetext{https://exoplanetarchive.ipac.caltech.edu/}
\subsection{Reduced formation conditions and disk evolution models}
\label{sec:disc5}

Our results can be compared with meteoritic compositions within the framework of disk evolution models to improve understanding of the factors driving the diverse Fe compositions of rocky bodies (Fig. \ref{fig:fig7}).
The timing and location of reduced (oxygen-poor) formation environments in the Solar System are not well quantified. The most reduced parent bodies are inferred to originate from NC reservoirs, formed within $\sim2$ Ma after CAIs, close to the Sun and at temperatures high enough to partially melt \citep{budde2016tungsten}. Although evidence of the early Solar System gas compositions is not well preserved, recent observations of protoplanetary disks supported by disk chemistry models show that C/O in the gas evolves dynamically over time \citep{mah2023close}. The inward migration of icy pebbles plays a key role in this evolution, influencing the composition of condensed species, including the relative proportions of metallic iron, silicates, and oxides \citep{johansen2021, grossman2008redox}.

In models of solar-analogue protoplanetary disks, the absence of significant gaps, such as those lacking pressure traps or large planetary bodies, ensures a steady inward drift of icy pebbles, maintaining a supply of oxygen to the inner disk. This promotes oxidized conditions in the gas phase for several million years, even in moderately viscous disks ($\alpha = 10^{3}$) \citep{mah2023close}. Consequently, planets forming inside the snowline under these conditions are likely to form in oxidized environments. However, in disks with gaps, reduced environments may develop closer to the star if water vapor is efficiently accreted onto the star without being replenished \citep{mah2024mind}, particularly in high and moderately viscous disks ($\alpha < 0.01$). Reduced meteoritic parent bodies, including Mercury and the proto-Earth's building blocks, likely formed under such circumstances, with C/O approaching unity \citep{grossman2008redox}. Furthermore, the formation of Jupiter likely impeded the inward migration of icy pebbles, creating a brief period during which reduced bodies formed in the inner disk \citep{desch2018effect}.

The challenge of forming high-density planets ($\geq 60 \%$ Fe by mass) or super-Mercuries solely through giant impacts \citep{cambioni2025can} has led to suggestions that rapid accretion of Fe-rich pebbles in inner disks also forms larger metal-rich planets \citep{johansen2022}. \cite{adibekyan2021compositional} suggested that high-density planets around FGK stars likely formed after the first few million years, in regions where water from within the snowline had already accreted onto the star \citep{mah2023forming}. Such conditions could have created a temporal window for the rapid formation of large, reduced, or Fe-rich bodies. Additional processes, such as the Curie temperature effect and Fe pebble clustering, likely influence the conditions and locations of super-Mercury formation \citep{bogdan2023curie}. Other mechanisms, such as photoevaporation, have also been proposed to explain the formation of dense exoplanets, with multiple factors likely acting in combination \citep{brinkman2024revisiting}.

\section{Conclusion} \label{sec:conclusion}
 Motivated by the need to understand Fe-rich planetary building blocks, this study systematically explores the interior compositions and structures of $\sim$200 km diameter bodies, starting from a range of bulk compositions (CI, H, EH, and CBa) and redox states ($-8 < \Delta\text{IW} < -3$). Differentiation models suggest that the core mass fraction does not vary significantly under moderate to highly reducing conditions ($\Delta\text{IW} < -3$). The results also indicate that Si dominates as a light core component under highly reduced conditions, whereas S, C, and FeO become significant under more oxidizing conditions. By lowering melting points, S-rich cores could support the generation of a dynamo, and sulfide liquids could also exsolve and be transported to mantles.  Additionally, carbon could exsolve from cores to form graphitic layers across a wide redox and compositional range. This study provides a framework to interpret measurements to be performed by asteroid rendezvous missions such as NASA's Psyche mission. These results may also provide insight into an emerging population of dense rocky exoplanets that, like reduced meteoritic parent bodies, likely formed within a restricted time window. During this period, the disk structure may have limited oxygen replenishment by preventing the inward migration of icy pebbles. A period of early reduced formation in the Solar System could also have contributed to the dearth of mantle silicates in the meteoritic record.

\section*{Data Availability} 
Composition data are available in Zenodo at 
\href{https://doi.org/10.5281/zenodo.17567495}{10.5281/zenodo.17567495}.

\begin{acknowledgements}
This work was supported by the Center for Matter at Atomic Pressures (CMAP), a National Science Foundation (NSF) Physics Frontiers Center, under Award PHY2020249 and by the National Aeronautics and Space Administration under Agreement No. 80NSSC21K0593 for the program “Alien Earths”. E.S.S. was supported by Marie Curie Postdoctoral Fellowship ``SuChaMa'' and ERC Starting Grant ``VenusVolAtmos''. This material is based upon work supported by the Department of Energy [National Nuclear Security Administration] University of Rochester “National Inertial Confinement Fusion Program” under Award Number DE-NA0004144.
\end{acknowledgements}

\bibliographystyle{aasjournal}
\bibliography{bibpsy}

\begin{appendix}
\section{Metal-silicate partition coefficients of the major elements}
\begin{figure}[htb!]
    \centering
    \includegraphics[width=0.5\textwidth, height=0.5\textheight, keepaspectratio]{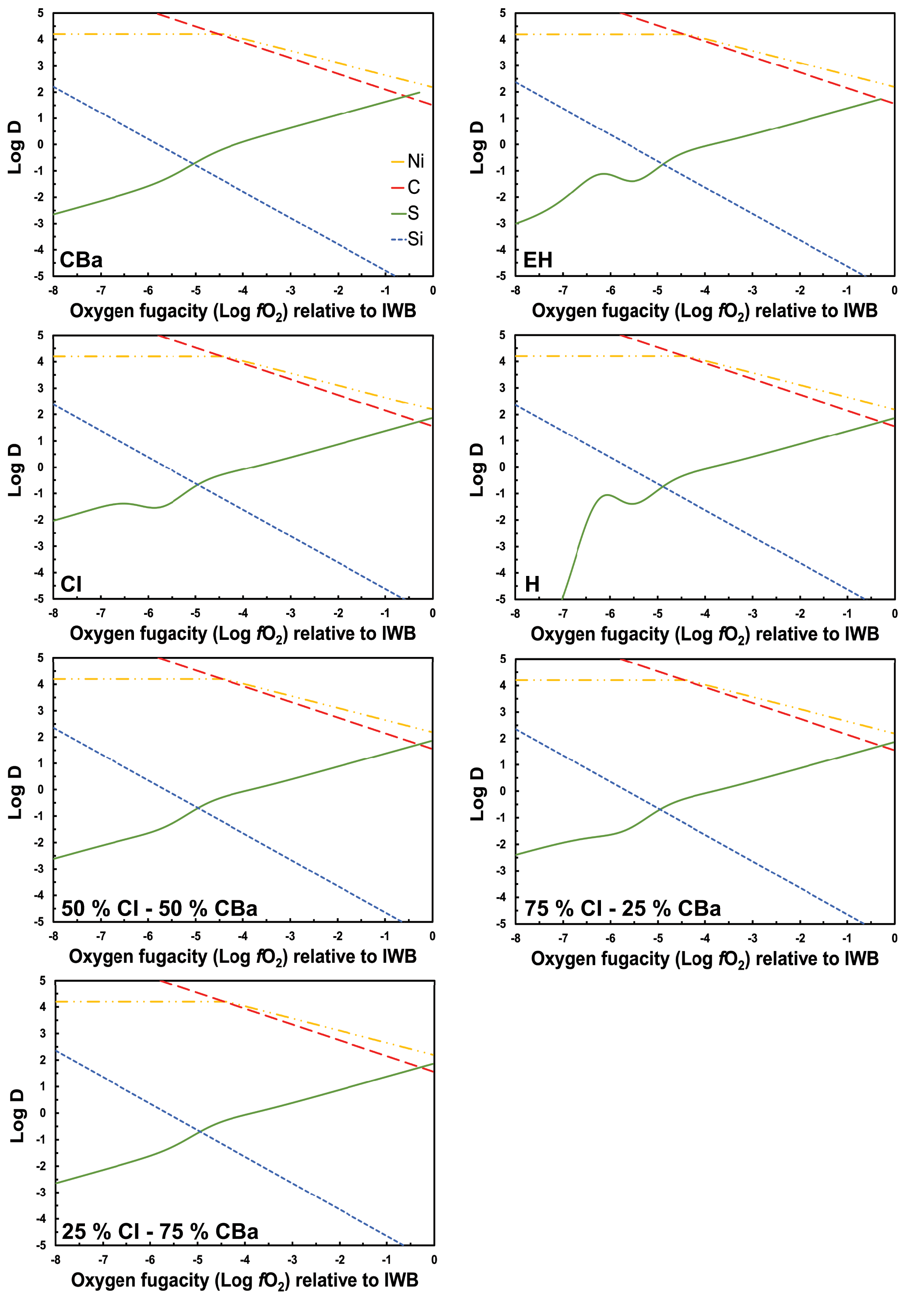}
    \caption{The metal-silicate partition coefficients, $Log D^\text{met-sil}_{i}$ plotted for key elements C, Si, S, and Ni as a function of $f\text{O}_2$, calculated at 1900 K and 0.1 GPa for the initial bulk compositions considered in this study. $Log D^\text{met-sil}_{i}$ functions are based on literature studies \citep{boujibar2014,steenstra2016, chi2014, Fischer2015, steenstra2020a}.} 
    \label{fig:A1}
\end{figure}

\section{Mantle compositions}
\begin{figure}[htb!]
    \centering
    \includegraphics[width=0.3\textwidth, height=0.4\textheight, keepaspectratio]{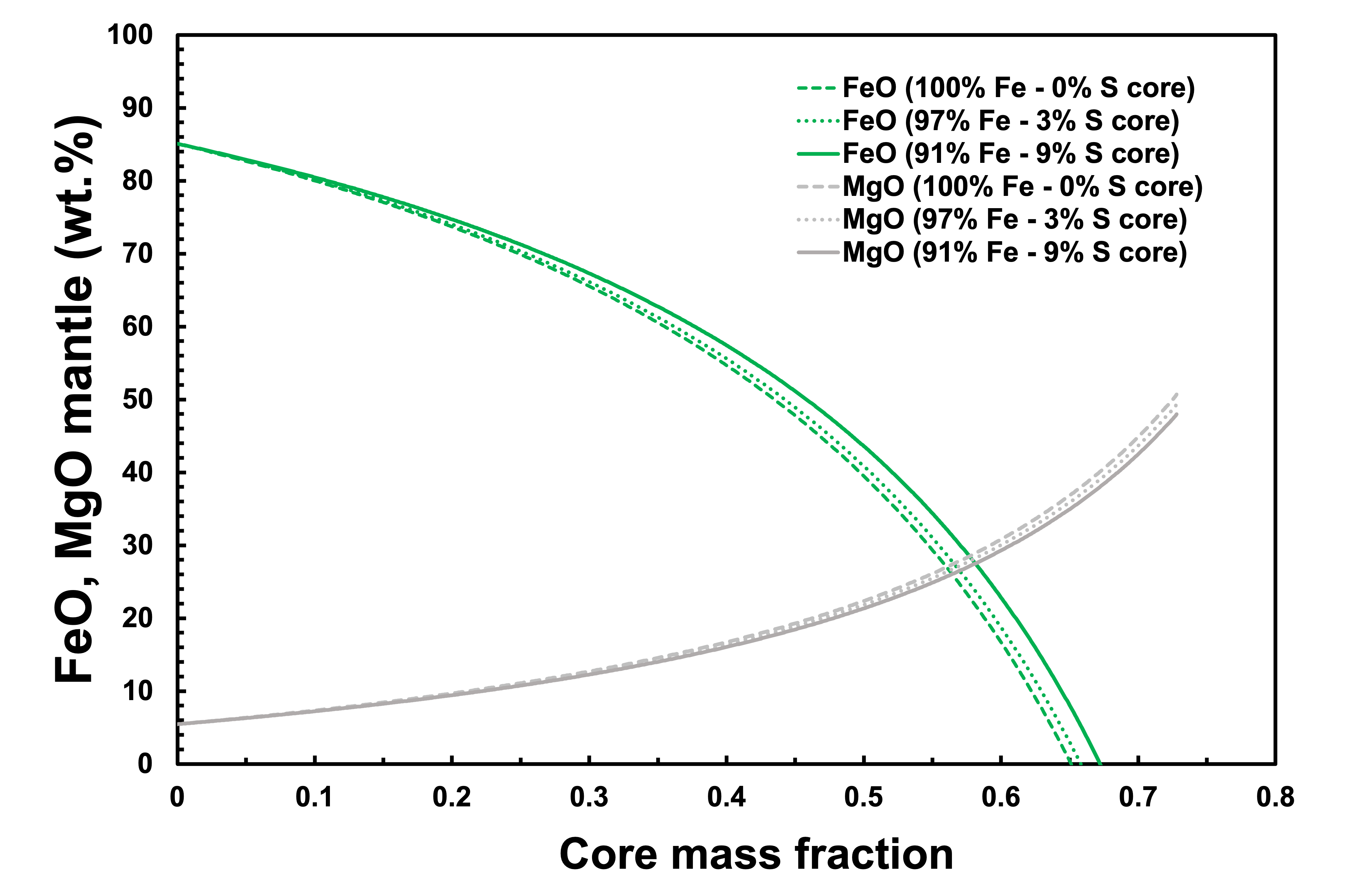}
    \caption{Major oxide mantle compositions computed using the methodology of \cite{palme2014}} 
    \label{fig:A2}
\end{figure}

\end{appendix}

\end{document}